\documentclass[12pt]{iopart}

\usepackage{cite}

\begin{document}

\title{HAM and HPM: Another attack to reason}
\author{Francisco M. Fern\'{a}ndez}
\address{INIFTA (UNLP, CCT La Plata-CONICET), Divisi\'on Qu\'imica Te\'orica,
Blvd. 113 S/N,  Sucursal 4, Casilla de Correo 16, 1900 La Plata,
Argentina}\ead{fernande@quimica.unlp.edu.ar}

\maketitle

\begin{abstract}
We discuss two recent applications of homotopy analysis method and homotopy
perturbation method and conclude that the results are completely useless from
both mathematical and physical points of view.
\end{abstract}

In two papers appeared recently in this journal El--Wakil and Abdou\cite
{EA10} and Abdou\cite{A10} proposed the application of homotopy analysis
method (HAM) and homotopy perturbation method (HPM), respectively, to
several nonlinear equations of supposed interest in fluid mechanics, plasma
physics, optical fibres, biology, solid state physics, chemical kinematics,
chemical physics, geochemistry\cite{EA10}, particle vibration in lattices,
currents in electrical networks, pulses in biological chains, etc.\cite{A10}.

These two papers are examples of the new physics and mathematical physics
that has lately spread among many journals. In what follows we discuss them.

El--Wakil and Abdou\cite{EA10} stated that ``Being different from the
perturbation technique, the HAM does not need any small parameter to
function''. However, if we look at the paper we realize that they expanded a
solution to the nonlinear equation in the way that is common to perturbation
theory: $u(x,t)=u_{0}(x,t)+u_{1}(x,t)p+\ldots +u_{n}(x,t)p^{n}+\ldots $ and
set the dummy perturbation parameter $p$ equal to unity at the end of the
calculation. What is $p$ if not a small parameter?. Most curiously, in the
other paper Abdou\cite{A10} stated that ``Using the homotopy technique in
topology, a homotopy is constructed with an embedding parameter $p\in (0,1)$
which is considered as a small parameter''. It seems that a dummy
perturbation parameter that plays exactly the same role in both approaches
is a small parameter in HPM\cite{A10} but it is not in HAM\cite{EA10}.

El--Wakil and Abdou\cite{EA10} studied the following system of coupled
nonlinear equations:
\begin{eqnarray}
\frac{\partial u}{\partial t} &=&u(1-u-v)+\frac{\partial ^{2}u}{\partial
x^{2}}  \nonumber \\
\frac{\partial v}{\partial t} &=&\frac{\partial ^{2}v}{\partial x^{2}}-uv
\label{eq:coupled_u,v}
\end{eqnarray}
with the initial conditions
\begin{eqnarray}
u(x,0) &=&\frac{e^{-kx}}{\left( 1+e^{-kx/2}\right) ^{2}}  \nonumber \\
v(x,0) &=&\frac{1}{1+e^{-kx/2}}  \label{eq:Ini_cond_u,v}
\end{eqnarray}
where $k$ is a constant. According to the authors ``The above system of two
coupled nonlinear equations of reaction diffusion type arise in chemical
reactions or in ecology, and other fields of physics''. It is remarkable
that they are able to solve problems in so many fields of scientific
interest with such simple ubiquitous system of equations. Unfortunately, the
authors did not give any reference about such applications.

They applied HAM and derived a most cumbersome approximate solution that
they displayed in their Appendix A. If you look at it you realize that it is
merely the set of Taylor expansions
\begin{eqnarray}
u(x,t) &=&\sum_{j=0}u_{j}(x)t^{j}  \nonumber \\
v(x,t) &=&\sum_{j=0}v_{j}(x)t^{j}  \label{eq:Taylor_u,v}
\end{eqnarray}
through third order. Obviously, if you substitute these equations into
equation (\ref{eq:coupled_u,v}) you easily obtain the coefficients $u_{j}(x)$
and $v_{j}(x) $ that the authors so laboriously derived by means of HAM.

According to El--Wakil and Abdou\cite{EA10} the exact solutions are
\begin{eqnarray}
u(x,t) &=&\frac{e^{kz}}{\left( 1+e^{kz/2}\right) ^{2}}  \nonumber \\
v(x,t) &=&\frac{1}{1+e^{kz/2}}  \label{eq:exact_u,v_1}
\end{eqnarray}
where $z=x+xt$. However, the reader may easily verify that they do not
satisfy the nonlinear equations (\ref{eq:coupled_u,v}). The authors stated
that those exact solutions come from another paper\cite{EA07} where one
reads
\begin{eqnarray}
u(z) &=&\frac{e^{kz}}{\left( 1+e^{kz/2}\right) ^{2}}  \nonumber \\
v(z) &=&\frac{1}{1+e^{z/2}}  \label{eq:exact_u,v_2}
\end{eqnarray}
and $z=x+ct$. Unfortunately, the second equation does not satisfy the
corresponding initial condition and, therefore, it cannot be a solution. It
seems that the system of nonlinear equations (\ref{eq:coupled_u,v}) exhibits
an exact solution but the authors failed to provide it in two papers were
they treated the same problem with two different approaches\cite{EA10,EA07}.

Abdou\cite{EA10} discussed the differential difference equations (DDEs)
\begin{equation}
\frac{\partial u_{n}}{\partial t}=\left( 1+\alpha u_{n}+\beta
u_{n}^{2}\right) \left( u_{n+1}-u_{n-1}\right)   \label{eq:DDEs}
\end{equation}
where $\alpha $ and $\beta $ are constants. The author did not say it
explicitly but it appears that $n$ is an integer. According the author ``The
hybrid nonlinear difference equation (\ref{eq:DDEs}) describes the
discretization of the Korteg--de Vries (KdV) and modified KdV equations with
initial condition''
\begin{equation}
u_{n,0}=a_{0}-\frac{1}{\alpha }\left( \alpha a_{0}+2\right) \tanh
^{2}(k)\tanh ^{2}(kn+c)  \label{eq:DDE's_ini}
\end{equation}
``whose exact solution reads as''
\begin{eqnarray}
u_{n}(t) &=&a_{0}-\frac{1}{\alpha }\left( \alpha a_{0}+2\right) \tanh
^{2}(k)\tanh ^{2}  \nonumber \\
&&\times \left[ kn+(\alpha a_{0}+2)^{2}\tanh (k)\mathrm{sech}^{2}(k)\frac{t}{%
2}+c\right]   \label{eq:DDEs_exact}
\end{eqnarray}
They applied the HPM and obtained just the first term of the Taylor
expansion of $u_{n}(t)$ about $t=0$. The author also treated other DDEs with
known exact solutions. Since they are simpler than the one discussed above
he obtained some more terms of the Taylor series.

Summarizing: El--Wakil and Abdou\cite{EA10} and Abdou\cite{A10} applied HAM
and HPM, respectively, and obtained Taylor series of differential equations
with known exact solutions. Such equations are merely tailor--made toy
problems with the purpose of applying those methods. One easily obtains the
time series by straightforward expansion of the equations about $t=0$. The
resulting approximate solutions are valid for short times, which are
typically of little or no physical interest at all. Although the authors
mentioned that the models appear in every branch of physics, chemistry, and
engineering, it is clear that they do not apply to actual problems of
interest.

Variational and perturbation approaches (VAPA) like HPM, HAM, Adomian
decomposition method (ADM), variational iteration method (VIM), etc. have
produced many pseudo scientific works of the poorest quality. We have
denounced this unhappy situation in several articles\cite
{F07,F08b,F08c,F08e,F08d,F08f,F09a,F09b,F09c,F09d,F09e,F09f,F09g} but
apparently nobody seems to care. We suggest the reader to have a look at
some of the articles cited by El--Wakil and Abdou\cite{EA10} and Abdou\cite
{A10} in order to have an idea of what we are talking about. There is a sort
of endogamic refereeing process that spreads VAPA like vermin and more
journals are being contaminated every day.


\begin{thebibliography}{99}
\bibitem{EA10}  El-Wakil S A and Abdou M A 2010 \textit{Phys. Scr.} \textbf{%
81} 015001 (8 pp.).

\bibitem{A10}  Abdou M A 2010 \textit{Phys. Scr.} \textbf{81} 015003 (8 pp.).

\bibitem{EA07}  El-Wakil S A and Abdou M A 2007 \textit{Chaos, Solitons \&
Fractals} \textbf{33} 513.

\bibitem{F07}  Fern\'{a}ndez F M, Perturbation Theory for Population
Dynamics, arXiv:0712.3376v1

\bibitem{F08b}  Fern\'{a}ndez F M, On Some Perturbation Approaches to
Population Dynamics, arXiv:0806.0263v1

\bibitem{F08c}  Fern\'{a}ndez F M, On the application of
homotopy-perturbation and Adomian decomposition methods to the linear and
nonlinear Schr\"{o}dinger equations, arXiv:0808.1515v1

\bibitem{F08e}  Fern\'{a}ndez F M, On the application of homotopy
perturbation method to differential equations, arXiv:0808.2078v2

\bibitem{F08d}  Fern\'{a}ndez F M, On the application of the variational
iteration method to a prey and predator model with variable coefficients,
arXiv.0808.1875v2

\bibitem{F08f}  Fern\'{a}ndez F M, Homotopy perturbation method: when
infinity equals five, 0810.3318v1

\bibitem{F09a}  Fern\'{a}ndez F M 2009 \textit{Phys. Scr.} \textbf{79}
055003 (2pp.).

\bibitem{F09b}  Fern\'{a}ndez F M, Amazing variational approach to chemical
reactions, arXiv:0906.0950v1 [physics.chem-ph]

\bibitem{F09c}  Fern\'{a}ndez F M, On the homotopy perturbation method for
Boussinesq-like equations, arXiv:0907.4481v1 [math-ph]

\bibitem{F09d}  Fern\'{a}ndez F M, Perturbation approaches and Taylor
series, arXiv:0910.0149v1 [math-ph]

\bibitem{F09e}  Fern\'{a}ndez F M, On a simple approach to nonlinear
oscillators, arXiv:0910.0600v1 [math-ph]

\bibitem{F09f}  Fern\'{a}ndez F M 2009 \textit{Appl. Math. Comput.} \textbf{%
215} 168.

\bibitem{F09g}  Fern\'{a}ndez F M, Nonsensical models for quantum dots,
arXiv:0911.3311v1 [math-ph]
\end{thebibliography}
\end{document}